\documentclass[12pt,reqno]{amsart}
\usepackage{amssymb}
\usepackage[usenames]{color}
\usepackage{comment}
\usepackage[shortlabels]{enumitem}
  \setlist[enumerate,1]{leftmargin=25pt}
  \setlist[itemize,1]{leftmargin=20pt}
  \setlist[description,1]{leftmargin=15pt}
\usepackage{qcircuit}
\usepackage[leftmargin=15pt,vskip=8pt]{quoting}
\usepackage{setspace}
\usepackage{url}
\usepackage{hyperref}
\newtheorem{thm}{Theorem}
\newtheorem{prop}[thm]{Proposition}

\newtheorem{cor}[thm]{Corollary}

\theoremstyle{definition}
\newtheorem{df}[thm]{Definition}
\newtheorem{conv}[thm]{Convention}
\newtheorem{notat}[thm]{Notation}

\theoremstyle{remark}

\newtheorem{rmk}[thm]{Remark}

\theoremstyle{definition}

\newenvironment{lquote}
  {\list{}{\leftmargin=1.5em\rightmargin=1em}\item[]}%
  {\endlist}

\newcommand{\ket}[1]{\ensuremath{|#1\rangle}}

\renewcommand\phi{\varphi}

\newcommand\qef{\hfill$\triangleleft$} 

\newcommand\x{\times}

\newenvironment{ls}{\begin{itemize}}{\end{itemize}}
\newenvironment{lsnum}{\begin{enumerate}}{\end{enumerate}}
\newenvironment{pf}{\begin{proof}}{\end{proof}}

\newcommand{\bbb}[1]{\ensuremath{\mathbb {#1}}}

\newcommand{\emp}{\varnothing}
\renewcommand{\phi}{\varphi}
\newcommand{\sq}[1]{\ensuremath{\langle#1\rangle}}

\newcommand{\notarrow}{\kern .42em\not\kern -.42em\longrightarrow}

\newcommand{\bl}{\ensuremath{\{0,1\}}}
\newcommand{\ql}{\ensuremath{Q}}
\newcommand{\exx}[1]{\ensuremath{\mathrm{Exit}(#1)}}

\newcommand{\noprint}[1]{\relax}

\title{Circuits: An  abstract viewpoint}
\author{Andreas Blass}
\address{Mathematics Department\\
University of Michigan\\
Ann Arbor, MI 48109--1043, U.S.A.}
\email{ablass@umich.edu}
\author{Yuri Gurevich}
\address{Computer Science and Engineering\\
University of Michigan\\
Ann Arbor, MI  48109-2121, U.S.A}
\email{gurevich@umich.edu}

\begin{document}
\newpage
\maketitle

\begin{abstract}
Our primary purpose is to isolate the abstract, mathematical properties of circuits --- both classical Boolean circuits and quantum circuits --- that are essential for their computational interpretation.  A secondary purpose is to clarify the similarities
and differences between the classical and quantum situations.

The general philosophy in this note is to include the mathematically essential aspects of circuits but to omit any of the additional structures that are usually included for convenience.  We shall, however, retain the assumption that circuits are finite; this assumption does no harm to the applicability of our approach and is necessary for some of our work.
\end{abstract}

\begin{quote}\raggedleft\small\it
One of the endearing things about mathematicians is the\\ extent to which they will go to avoid doing any real work.\\[1ex]
--- Matthew Pordage
\end{quote}

\section{Introduction} 
\label{sec:intro}

%

As we worked on Circuit Pedantry \cite{G242}, we tried to figure out
the appropriate level of abstraction for Boolean and quantum
circuits. There is a natural tendency in the sciences, but especially
in mathematics, to abstract away as many details as possible. This
celebrated tendency is fruitful but it may also be fraught with
troubles of various kinds.

The story goes that Plato defined man as featherless biped,
abstracting from man's many other properties, but Diogenes plucked the
feathers from a cock and brought it to Plato saying: ``Here's your
man."

A more recent example is Cantor's definition of sets. Cantor imposed
no restrictions on what the elements of a set can be or how they are
collected into a whole. The notion of set simplified mathematical
analysis, enabled the development of logic and topology, etc. But it
also led to paradoxes.

Here is a more pedestrian example which is closer to Circuit
Pedantry. A finite matrix can be defined as an indexed set
$\{e_p: p\in R\x C\}$ where $R$ and $C$ are finite sets indexing rows
and columns. Normally the rows are linearly ordered and so are the
columns. But the notion of finite matrix makes perfect sense without
those orderings. That is, until you want to draw a matrix on a
whiteboard or on paper.

Still, there are advantages in dealing with sets and indexed sets
rather than linearly ordered sets. Think of relational databases where
relational tuples are unordered, which simplifies theory \cite{GUW}
and improves practice \cite{DD} by eliminating a most important source
of implementation dependence.

The set-based approach supports the most expressive polynomial-time
computation model in the literature where machines do not distinguish
between isomorphic structures \cite{G120,G150,DRR}. Some
complexity-theoretic advantages of the set-based approach are
demonstrated in \cite{G139}.

In Circuit Pedantry, we restrained our own tendency to abstract and
adopted the traditional approach. In that approach, the input nodes
are ordered in an arbitrary way and --- in the case of quantum
circuits (and balanced Boolean circuits) --- one fixes a bijection
between incoming and outgoing edges for every gate. As a result, there
are definite timelines from the input nodes to output nodes. This
allowed us to use traditional circuit diagrams as in, for example,
\cite{NC} and hopefully to make that paper more readable.

But before we restrained our tendency to abstract, we indulged it for
a little while. There is elegance and mathematical utility in the
more abstract view. We are using the abstract view in our forthcoming
paper on Quantum Circuits with Classical Channels \cite{G245} and we
illustrate it here.

Our primary purpose in the present paper is to isolate the abstract,
mathematical properties of circuits --- both classical Boolean
circuits and quantum circuits --- that are essential for their
computational interpretation.  A secondary purpose is to clarify the
similarities and differences between the classical and quantum
situations.

Our general philosophy in this note is to include the mathematically
essential aspects of circuits but to omit any of the additional
structures that are usually included for convenience.  We shall,
however, retain the assumption, satisfied in theoretical as well as
applied work, that circuits are finite.  This assumption does no harm
to the applicability of our approach and is necessary for some of our
work.

In the rest of this introduction, we describe how we want to view circuits. Precise details will be given in later sections.

Inputs and outputs, whether of a whole circuit or of a single gate, will be families (of Boolean values or of quantum states, usually qubits and possibly entangled) indexed by some finite sets.  It is customary to index inputs and outputs by natural numbers, thereby imposing a linear ordering on the inputs and another linear ordering on the outputs. When the number of inputs equals the number of outputs, we thereby obtain a particular bijection between the inputs and outputs. Although such bijections are useful for drawing circuits, we shall see that none of this customary extra structure --- numerical indexes, linear orders, particular bijections --- is essential for mathematical purposes; indeed none of this structure will appear in the formal development below. Any gate $G$ will have a finite set $\iota_G$ of input labels and a finite set $o_G$ of output labels, but there will be no additional structure or assumptions on these sets. (We use the Greek letters iota and omicron for input and output of gates, in order to keep $i$ available for other uses.)

In the case of Boolean gates, the whole input will be an $\iota_G$-indexed family of Boolean values, i.e., an element of $\bl^{\iota_G}$, and the output will be an element of $\bl^{o_G}$.  In the case of quantum gates, the input and output will be vectors in $\ql^{\otimes \iota_G}$ and $\ql^{\otimes o_G}$, respectively, where $\ql$ is our basic Hilbert space, usually $\bbb C^2$, the state space for a qubit.

Similarly, the circuit as a whole will have input nodes indexed by a
finite set $I$ and output nodes indexed by another finite set $O$,
with no additional structure or assumptions. The input to such a
circuit will be in $\bl^I$ in the Boolean case or $\ql^{\otimes I}$ in
the quantum case. The output will be in $\bl^O$ in the Boolean case or
$\ql^{\otimes O}$ in the quantum case.

The connections between gates, inputs, and outputs will also be
described in what we believe to be the simplest reasonable
way. Wherever a value (Boolean or quantum) is needed, there will be a
pointer to a provider for that value. A value is needed at each input
position of a gate and at each output node of the whole
circuit. Potential providers for these values are the circuit's input
nodes and the gates' output positions. We call the places where a
value is needed ``consumers'' and the places where a value can be
obtained ``producers''. So the wiring of our circuits will be given by
a ``provider'' function $\pi$ from consumers to producers, giving for
each consumer $c$ a producer $\pi(c)$ expected to supply the value
needed by $c$.

\section{Preliminaries}

Throughout this paper we shall need to work with families indexed by
arbitrary finite sets, in contexts where indexing by natural numbers
is more common and provides a specific ordering for the elements of
the family.  This preliminary section is devoted to describing how our
more general sort of indexing works and fixing our notation for it.

\begin{conv}
Throughout this paper, index sets are assumed to be finite.
\end{conv}

An $I$-indexed \emph{family} is a function $x$ with domain $I$. The
usual notations for the value of $x$ at $i$ are $x(i)$ and $x_i$. The
family itself is usually written \sq{x_i:i\in I} or $\sq{x_i}_{i\in
  I}$.  The $x_i$'s are called the \emph{elements} or
\emph{components} of the family.

When the index set $I$ is $\{1,2,\dots,n\}$, one may write such
an indexed family as $\sq{x_1,x_2,\dots,x_n}$ and call it an
$n$-\emph{tuple}.

We use the notation $\bigsqcup_{i\in I}A_i$ for the \emph{disjoint
  union} of a family \sq{A_i:i\in I} of sets, defined as
\[
\bigsqcup_{i\in I}A_i=\{\sq{i,a}:i\in I\text{ and }a\in A_i\}.
\]
In other words, we replace all the sets $A_i$ by pairwise disjoint,
bijective copies, $\{i\}\times A_i$, and then we take the union of
those copies. If the $A_i$'s are themselves pairwise disjoint, then we
could have just taken their union without copying, and we may tacitly
identify that union with the official disjoint union defined
above. Even when the $A_i$'s are not disjoint, we may tacitly identify
elements $a$ of $A_i$ with the corresponding elements \sq{i,a} of
$\bigsqcup_{i\in I}A_i$, relying on the context to provide the correct
$i$.

When the index set $I$ is $\{1,2,\dots,n\}$, one may write the
disjoint union as
\[
\bigsqcup_{i\in \{1,2,\dots,n\}}A_i=\bigsqcup_{i=1}^nA_i=
A_1\sqcup A_2\sqcup\dots\sqcup A_n.
\]
In particular, we have the binary operation $\sqcup$ as in
$A_1\sqcup A_2$.

The \emph{(Cartesian) product} of a family \sq{A_i:i\in I} is defined
as the collection of those $I$-indexed families whose elements are
taken from the corresponding sets $A_i$.  That is,
\[
\prod_{i\in I}A_i=\{\sq{x_i:i\in I}:(\forall i\in I)\,x_i\in A_i\}.
\]

As in the case of disjoint unions, alternative notations may be used
when $I=\{1,2,\dots,n\}$, namely
\[
\prod_{i\in \{1,2,\dots,n\}}A_i=\prod_{i=1}^nA_i=
A_1\times A_2\times\dots\times A_n.
\]
In particular, we have the binary operation $\times$ as in
$A_1\times A_2$.

Similar conventions apply to the tensor product of vector spaces. The
tensor product of an indexed family \sq{V_i:i\in I} of vector spaces
can be defined as the vector space generated by the elements of
$\prod_{i\in I}V_i$, considered as formal symbols, modulo the
relations that make the generators linear functions of each component
when the other components are held fixed.

The precise definitions will be given in a moment, but let us first
give an orienting example with $I=\{1,2,3\}$. A typical generator
\sq{x_1,x_2,x_3} would, in the context of tensor products, often be
written as $x_1\otimes x_2\otimes x_3$, and a fairly typical relation
would be the distributivity equation
\[
(x_1\otimes p\otimes x_3)+(x_1\otimes q\otimes x_3)=
x_1\otimes(p+q)\otimes x_3.
\]

To describe these relations in more detail and in full generality, it
is convenient to introduce a bit of notation. If \sq{x_i:i\in I} is an
indexed family, if $j\in I$, and if $q$ is an arbitrary entity, then
we write ``\sq{x_i:i\in I} but $j\mapsto q$'' for the $i$-indexed
family \sq{x'_i:i\in I} where $x'_i=x_i$ for all $i\neq j$ but
$x'_j=q$. That is, we modify the original famiiy \sq{x_i:i\in I} by
changing the $j$-component to $q$. Then the linearity relations for
the tensor product are, first, for all $j\in I$ and all $p,q\in V_j$,
\begin{multline*}
\big(\sq{x_i:i\in I}\text{ but }j\mapsto p\big)+
\big(\sq{x_i:i\in I}\text{ but }j\mapsto  q\big)=\\
=\sq{x_i:i\in I}\text{ but }j\mapsto p+q,
\end{multline*}
and second, for all $j\in I$, all $q\in V_j$, and all scalars
$\lambda$,
\[
\lambda\big(\sq{x_i:i\in I}\text{ but }j\mapsto q\big)=
\sq{x_i:i\in I}\text{ but }j\mapsto \lambda q.
\]
We use the notation $\bigotimes_{i\in I}V_i$ for this tensor
product. As before, when $I=\{1,2,\dots,n\}$, we have the alternative
notations
\[
\bigotimes_{i\in \{1,2,\dots,n\}}A_i=\bigotimes_{i=1}^nA_i=
A_1\otimes A_2\otimes\dots\otimes A_n,
\]
and we have the binary operation $\otimes$ as in $A_1\otimes A_2$.

When people work with numerical index sets and with the binary
operations $\sqcup$, $\times$, and $\otimes$, they make extensive (but
often tacit) use of the commutative and associative laws (up to
canonical isomorphism) for these operations, and the main effect of
these laws is to render the numerical indexing irrelevant. In our
general indexed context, these laws take on quite different forms.
Commutativity in the numerical-indexed context allows one to change
the order of the operands, but our operands don't come with an
order. Associativity in the numerical context allows one to regard
operations on three or more operands as built up from binary
operations in various ways, but we have defined the $n$-ary and in
fact $I$-ary operations directly, not in terms of binary ones. We list
below the more general laws governing our more general operations. In
each case, the isomorphisms indicated by $\cong$ are obvious and
will be referred to as canonical.  We leave the routine verifications
to the reader.

Suppose $f:I\to J$ is a bijection. Then any $J$-indexed family, being
a function with domain $J$, can be composed with $f$ to produce an
$I$-indexed family, with the same components but differently
indexed. In symbols, composition with $f$ transforms \sq{x_j:j\in J}
into \sq{x_{f(i)}:i\in I}. Then we have, for any families \sq{A_j:j\in
  J} of sets and \sq{V_j:j\in J} of vector spaces,
\begin{align*}
  \bigsqcup_{i\in I} A_{f(i)}&\cong\bigsqcup_{j\in J}A_j\\
  \prod_{i\in I} A_{f(i)}&\cong\prod_{j\in J}A_j\\
  \bigotimes_{i\in I} V_{f(i)}&\cong\bigotimes_{j\in J}V_j.
\end{align*}
That is, up to canonical isomorphisms, re-indexing doesn't change
disjoint unions, Cartesian products, and tensor products.

Now suppose \sq{J_i:i\in I} is an $I$-indexed family of index sets
$J_i$, and let $K$ be the disjoint union of all the $J_i$ (remember
that elements of $K$ have the form \sq{i,j} with $i\in I$ and
$j\in J_i$). Then we have, for any $K$-indexed families
\sq{A_{\sq{i,j}}: \sq{i,j}\in K} of sets and \sq{V_{\sq{i,j}}:
  \sq{i,j}\in K} of vector spaces,
\begin{align*}
\bigsqcup_{\sq{i,j}\in K}A_{\sq{i,j}}&\cong
\bigsqcup_{i\in I}\bigsqcup_{j\in J_i}A_{\sq{i,j}}\\
\prod_{\sq{i,j}\in K}A_{\sq{i,j}}&\cong
\prod_{i\in I}\prod_{j\in J_i}A_{\sq{i,j}}\\
\bigotimes_{\sq{i,j}\in K}V_{\sq{i,j}}&\cong
\bigotimes_{i\in I}\bigotimes_{j\in J_i}V_{\sq{i,j}}.
\end{align*}

We shall sometimes simplify notation by omitting mention of the
canonical bijections and isomorphisms above. For example, if
$X=A\times C$ and $Y=B\times D$, then we may identify $A\times
B\times C\times D$ with $X\times Y$, omitting mention of the canonical isomorphism arising from the bijection $f: I=\{1,2,3,4\}\to
J = \bigsqcup_{i\in\{1,2\}} \{1,2\}$ that sends $1,2,3,4$ to
$\sq{1,1},\sq{2,1},\sq{1,2},\sq{2,2}$, respectively.
In detail, let $A,B,C,D = A_{11}, A_{21}, A_{12}, A_{22}$. Then
\begin{align*}
A\times B\times C\times D
  &= A_{11} \times A_{21} \times A_{12}\times A_{22}
  = \prod_{i\in I}A_{f(i)} \\
  &\cong \prod_{j\in J} A_j \\
  &\cong \prod_{u\in \{1,2\}} \prod_{v\in\{1,2\}} A_{\sq{u,v}}
  = (A\x C)\x (B\x D).
\end{align*}

Consider two $I$-indexed families of sets \sq{A_i:i\in I} and
\sq{B_i:i\in I} and an $I$-indexed family of functions
$f_i:A_i\to B_i$.  These functions $f_i$ induce functions on disjoint
unions and Cartesian products
\[
\sq{i,a}\mapsto\sq{i,f_i(a)}:
\bigsqcup_{i\in I}A_i\to\bigsqcup_{i\in I}B_i
\]
and
\[
\sq{a_i:i\in I}\mapsto\sq{f_i(a_i):i\in I} :
\prod_{i\in I}A_i\to\prod_{i\in I}B_i.
\]

Similarly, linear transformations between vector spaces
$f_i:V_i\to W_i$ induce a linear transformation of the tensor
products, $\bigotimes_{i\in I}V_i\to\bigotimes_{i\in I}W_i$\,, sending
each generator \sq{x_i:i\in I} of the former space to the generator
\sq{f_i(x_i):i\in I} of the latter. It is easy to check, using the
linearity of the $f_i$'s, that this mapping of the generators respects
the defining relations of the tensor product and thus gives a
well-defined linear transformation of the tensor products.

\begin{rmk}
For category-minded readers, we mention that $\bigsqcup$ and $\prod$
are functors from $I$-indexed families of sets to sets. In fact, they
are the left and right adjoints, respectively, of the functor that
sends any set $X$ to the $I$-indexed family all of whose components
are $X$.  Similarly, $\bigotimes$ is a functor from $I$-indexed
families of vector spaces to vector spaces. The canonical isomorphisms
indicated earlier are natural isomorphisms in the category-theoretic
sense.
\end{rmk}

\section{Boolean Circuits}

\begin{df}
  A \emph{Boolean gate type} is a triple \sq{\iota,o,g} consisting of
  two finite sets $\iota$ and $o$ and a function
  $g:\bl^\iota\to\bl^o$. We call $\iota$ the set of input labels, $o$
  the set of output labels, and $g$ the function of the gate type.
\end{df}

In this section, all gate types under consideration will be Boolean, so we
omit ``Boolean'' and just call them gate types.

People often restrict the labels to be natural numbers. This makes it
easier to write elements of $\bl^\iota$ and $\bl^o$, but it has no
mathematical significance.

\begin{df}
A \emph{Boolean circuit} consists of
\begin{ls}
\item a finite set $I$ of \emph{input nodes},
\item a finite set $O$ of \emph{output nodes},
\item a finite set of  \emph{gates},
\item an assignment of a gate type $(\iota_G,o_G,g_G)$ to each gate
  $G$, and
\item a \emph{provider} function $\pi$ as described below.
\end{ls}
By \emph{producers} we mean input nodes and triples of the form
$(G,\text{out},l)$ where $G$ is a gate of the circuit and $l$ is one
of its output labels ($l\in o_G$). We call such a triple an
\emph{output port} of the gate $G$. By \emph{consumers} we mean output
nodes and triples of the form $(G,\text{in},l)$ where $G$ is a gate of
the circuit and $l$ is one of its input labels ($l\in \iota_G$). We
call such a triple an \emph{input port} of the gate $G$. Producers and
consumers are called \emph{nodes} of the circuit.

The provider function $\pi$ is a function from consumers to producers
subject to the following requirement. We say that a gate $G$ is a
\emph{direct prerequisite} for another gate $H$ and we write $G\prec
H$ if $\pi$ maps (at least) one of the input ports of $H$ to an output
port of $G$. We require that the relation $\prec$ be acyclic. \qef
\end{df}

To simplify terminology and notation, we shall sometimes refer to
providers of a gate when we mean providers of that gate's input
ports. Thus, $G\prec H$ if and only if some provider of $H$ is an
output port of $G$.

\begin{notat}           \label{pi-gates}
  When $G$ is a gate, we abbreviate
  $\{\pi(G,\text{in},l):l\in\iota_G\}$ as $\pi(G)$.
\end{notat}

In view of the assumption that our circuits are finite, the
requirement that $\prec$ be acyclic is equivalent to requiring that it
be a well-founded relation. We use the word \emph{prerequisite}
without ``direct'' and the notation $\prec^*$ for the transitive
closure of $\prec$; thus $\prec^*$ is a strict partial order.

The intuition behind the definition is as follows. Each gate $G$,
given the set $\iota_G$ of input labels, the set $o_G$ of output
labels, and the Boolean function $g_G$, reads an input in
$\bl^{\iota_G}$ from its input ports, applies $g_G$, and puts the
result in $\bl^{o_G}$ at its output ports. The inputs here, at the
input ports $x$ of $G$, are simply retrieved from the corresponding
nodes $\pi(x)$ as given by the provider function $\pi$. The gate $G$
consumes its inputs and produces its outputs; hence the ``producer''
and ``consumer'' terminology. The input nodes of the circuit, the
elements of $I$, can also provide inputs for gates, so they count as
producers. The output nodes can retrieve values computed by gates or
supplied in the input (as given by $\pi$) and exhibit them as the
result of the circuit's computation. This intuition is formalized in
the following theorem.

\begin{thm}     \label{b-thm}
Let a circuit be given along with an assignment of Boolean values to
its input nodes, i.e., an element $a$ of $\bl^I$. Then there is a unique
function $C$ assigning to each node $x$ of the circuit a Boolean
value $C(x)$ subject to the following requirements.
\begin{lsnum}
  \item For input nodes $x$, we have $C(x)=a(x)$.
\item For consumers $x$, we have $C(x)=C(\pi(x))$ (i.e., consumer nodes
  just retrieve bits from their providers).
\item For any gate $G$, its $o_G$-tuple of  outputs,
\[
l\mapsto C(G,\mathrm{out},l),
\]
is the result of applying its function $g_G$ to its $\iota_G$-tuple of
inputs
\[
m\mapsto C(G,\mathrm{in},m).
\]
\end{lsnum}
\end{thm}

\begin{pf}
Clause (2) reduces the problem to defining $C$ on producers. Rewriting
clause~(3) in terms of producers,
\[
(l\mapsto C(G,\text{out},l))=g_G(m\mapsto C(\pi(G,\text{in},m))),
\]
we find that this and clause~(1) constitute a definition of $C$ (on
producers) by recursion on the direct prerequisite relation
$\prec$. Since this relation is well-founded, the recursion has a
unique solution.
\end{pf}

According to the theorem, any $a\in\bl^I$ gives rise, via the function
$C$, to a uniquely defined element $b\in\bl^O$, namely the restriction
of $C$ to output nodes. In this way, the given circuit defines a
function $\bl^I\to\bl^O$, the function \emph{computed} by the
circuit.

\section{Balanced Boolean Circuits}

In preparation for the discussion of quantum circuits, we introduce a
special class of Boolean circuits, defined in \cite{G242} and designed
to be subject to some of the restrictions that become necessary when
one moves from the classical world to the quantum world.

\begin{df}
A Boolean circuit is \emph{balanced} if all of its gate functions
$g_G$ and its provider function $\pi$ are bijective.
\end{df}

Since bijective functions are invertible, bijectivity of all the gate
functions $g_G$ says that the circuit is composed entirely of
reversible gates.

Injectivity of the provider function $\pi$ means that each input bit and
each bit produced by a gate can be used (or output) only once. This
amounts to saying that the gates and inputs have no fan-out.

Surjectivity of $\pi$ means that input bits and bits produced by gates
must be used, either in computations by subsequent gates or as output
from the circuit. They cannot simply be discarded.  Intuitively, this
seems to be a mild requirement because, if a circuit did discard some
of its produced bits, then we could simply regard those bits as
additional output.  In other words, if $\pi$ were merely injective and
not bijective, we could enlarge $O$ and extend $\pi$ to map the new
elements of $O$ to those producers that were missing from the image of
$\pi$.

We record some immediate consequences of the definition.

First, if $G$ is a gate in a balanced circuit, then, since
$g_G:\bl^{\iota_G}\to\bl^{o_G}$ is a bijection, the index sets $\iota_G$ and
$o_G$ must have the same cardinality; each gate has equally many
input as output ports.\footnote{This observation would remain valid
  if each consumer $c$ received from its provider $\pi(c)$ not a bit but
  an element of some other, fixed alphabet $\Sigma$. But it would not
  be valid if the alphabet $\Sigma$ were allowed to be different for
  different $c$. For example, if a gate $G$ has hexadecimal inputs and
  binary outputs, then in order for $g_G$ to be bijective, $o_G$ must
  have four times as many elements as $\iota_G$.}

Summing that equality over all gates, we find that the total number of
gate input ports, which is the number of consumers except for the
circuit's output nodes, must equal the total number of gate output
ports, which is the number of producers except for the circuit's input
nodes.

But the provider function $\pi$ is also required to be a bijection, so the
number of consumers equals the number of producers, without the
exceptions.  Therefore, the exceptions must match, i.e., the circuit
has as many input nodes as output nodes: $|I|=|O|$.

\section{Quantum Gates and Circuits}

We turn now to the description of circuits for quantum
computation. For simplicity and to maintain similarity with the
Boolean case discussed in the preceding sections, we make two
assumptions about our circuits. First, we assume that the capacity of
each connection is a qubit, the quantum analog of a bit, rather than a
more complicated quantum system (which would be analogous to
transmitting more than one bit, or perhaps an element of some other
alphabet, in the Boolean case).  Second, we assume that each gate
represents a unitary operator; that is, we do not permit more
complicated\footnote{The general notion of quantum measurement, as
  defined in, for example \cite{NC}, allows measurements with only one
  possible outcome; such a measurement amounts to a unitary operator
  acting on the state.} measurements.  The second assumption is
eliminated in \cite{G245}.

Under these assumptions, quantum circuits differ from Boolean circuits
in the following ways. First, the no-cloning theorem means that at
most one consumer can use the output of any one producer, i.e., a
producer's output cannot be duplicated to supply multiple
consumers. Thus, the provider function $\pi$ of a quantum circuit is
necessarily one-to-one. Furthermore, just as in our earlier discussion
of Boolean circuits, we may assume that $\pi$ is surjective, i.e., that
whatever is produced is also consumed; we just treat any unconsumed
production as additional output. Thus, we may assume that the provider
function $\pi$ is bijective.

Second, the gate functions $g_G$ in a quantum circuit are not Boolean
functions but unitary transformations of Hilbert spaces. Specifically,
if $\iota_G$ and $o_G$ are, as before, the sets of input and output
labels, respectively, of $G$, then $g_G$ unitarily maps
$Q^{\otimes \iota_G}$ to $Q^{\otimes o_G}$, where $Q$ is the one-qubit
Hilbert space $Q=\bbb C^2$.  Since unitary transformations exist only
between Hilbert spaces of equal dimension, we conclude, just as in the
balanced Boolean case, that $|\iota_G|=|o_G|$ for every gate $G$ and
that therefore also $|I|=|O|$. In these respects, quantum circuits
look like balanced Boolean circuits.

Third, and most important, both for the utility of quantum computation
and for our work below, is entanglement. In the Boolean case, the
inputs to a gate were separate bits, obtained independently from the
appropriate providers. In the quantum case, it is usually not the case
that a gate's input qubits are independent. They may be entangled with
each other and also with other qubits that the gate in question does
not directly work with. This entanglement can be seen as the source of
the power of quantum computation; it is also the source of some of the
complexity in our formal development of the theory.

We now begin the formal development, interspersed with commentary to
clarify the underlying intentions.

\begin{notat}
  We use $Q$ to denote the qubit Hilbert space $\bbb C^2$.
\end{notat}

\begin{df}
  A \emph{quantum gate type} is a triple \sq{\iota,o,U} consisting of two
  finite sets $\iota$ and $o$ and a unitary transformation
  $U:Q^{\otimes \iota}\to Q^{\otimes o}$. We call $\iota$ the set of input
  labels, $o$ the set of output labels, and $U$ the operator of the
  gate type.
\end{df}

As mentioned above, unitarity of $U$ in this definition forces $\iota$
and $o$ to have the same cardinality. They need not, however, be the
same set, nor need there even be a canonical bijection between
them. In many pictures of quantum circuits, a particular bijection
would be implicit in the layout of the circuit on the page, but
neither the layout nor the bijection is canonical, and neither is
relevant in our abstract context.

\begin{df}
A \emph{quantum circuit} consists of
\begin{ls}
\item a finite set $I$ of \emph{input nodes},
\item a finite set $O$ of \emph{output nodes},
\item a finite set of \emph{gates},
\item an assignment of a gate type $(\iota_G,o_G,U_G)$ to each gate
  $G$, and
\item a bijective \emph{provider} function $\pi$ from consumers to
  producers such that the direct prerequisite relation $\prec$ and
  therefore also its transitive closure $\prec^*$ are acyclic.
\end{ls}
\end{df}

In the last clause of this definition, ``consumer'', ``producer'',
$\prec$, and $\prec^*$ are to be understood exactly as in the case of
Boolean circuits. Thus, the only difference between quantum circuits
and balanced Boolean circuits is that each gate $G$ has a unitary
operator $U_G:Q^{\otimes \iota_G}\to Q^{\otimes o_G}$ instead of a Boolean
bijection $g_G:\{0,1\}^{\iota_G}\to\{0,1\}^{o_G}$.  We also carry over
from the Boolean case Notation~\ref{pi-gates} and the terminology
``providers of a gate''.

The intuition behind the behavior of a quantum circuit is similar in
some respects to that for Boolean circuits but quite different in
other respects.

As before, a gate $G$ will obtain its input from its providers
and act on that input to produce its
output.
It is, however, important not to misinterpret ``retrieve'' in
our description of the Boolean case, ``inputs \dots\ are simply
retrieved from the corresponding provider nodes.'' ``Retrieve'' must not
mean ``copy'' here because quantum states, unlike classical bits,
cannot simply be copied. We should rather regard what is consumed at a
gate $G$ to be the same as (not a copy of) what is produced at its
provider nodes.

Furthermore, it can be misleading to speak of the state vector on
which a gate acts or the state vector that it produces. These states
will usually be entangled with other parts of the circuit that are not
directly involved with $G$. Even taking into account that the bits in
$\{0,1\}$ of the Boolean situation must be replaced by vectors in $Q$
in the quantum situation, we cannot expect to assign a state vector in
$Q$ to each node of the circuit;\footnote{We could assign a mixed
  state to each node by taking a suitable trace of the global
  state. The trace operation could, however, lose a great deal of
  information and could, in fact, ruin the usefulness of quantum
  computation. The reason is that tracing can destroy the entanglement
  on which quantum computation depends for its power.} we cannot
expect a direct analog of Theorem~\ref{b-thm}. Instead of keeping
track of separate bits at all the nodes, we must now keep track of the
evolution of a global quantum state.  Specifically, it makes good
sense to speak of the input state where the circuit's computation
begins, of the final state after the computation is complete, and of
various intermediate states, related to each other by the action of
the gates. The following definition serves to describe the contexts in
which such a global state makes sense.

\begin{df}
A \emph{stage} of a quantum circuit is a set $Z$ of gates closed under
direct prerequisites, i.e., if $x\in Z$ and $y\prec x$ then $y\in
Z$. The \emph{exits} of a stage $Z$ are those input nodes in $I$ and
output ports of gates $G$ in $Z$ that are not consumed in $Z$ (i.e.,
are not in $\pi(H)$ for any gate $H$ in $Z$).
We write
$\text{Exit}(Z)$ for the set of exits of a stage $Z$.
\end{df}

The formal notion of stage introduced in this definition is intended
to model the informal notion of a stage during a computation, that is,
a moment when some gates have already fired and the rest are still
waiting to fire. The set $Z$ consists of the gates that have already
fired, the ``past'' of the stage in question; the complementary set of
all gates not in $Z$ is the ``future'' of the stage. The requirement,
in the definition, that $Z$ be closed under $\prec$ formalizes the
idea that a gate cannot be fired until all its prerequisites have been
fired; firing a gate requires the availability of its input. Note that
closure under $\prec$ immediately implies closure under $\prec^*$.

The exits of a stage are those producers which have already produced
their outputs but have not yet had those outputs consumed. These
outputs constitute the information created (or supplied as input) in
the past and destined to be consumed in the future.

In terms of typical pictures of circuits, a stage $Z$ can be depicted
as a cut through the circuit, separating the gates already fired
(those in $Z$) from the rest of the gates, which still await firing in
the future. The circuit's input nodes in $I$ would be depicted as
being on the past side of the cut (where $Z$ is) while the output
nodes in $O$ are on the future side. The edges in the picture that
cross the cut are those whose past ends are in $Z$ (more precisely,
these ends are output ports of gates in $Z$) or $I$ and whose future
ends are not.  When people use such pictures, they often think in terms
of a global state associated to such a cut.

In our abstract picture, we don't directly refer to edges, but our
exits correspond to the past ends of the edges crossing the cut (and
their pre-images under $\pi$ correspond to the future ends of those
edges).

If one were to actually cut a circuit into a past circuit (input nodes
and gates in $Z$) and a future circuit (output nodes and gates not in
$Z$), then \exx Z would amount to providers for outputs of the past
fragment and to inputs for the future fragment. The terminology
``exit'' is intended to suggest the operation of these nodes in
producing output from the $Z$ fragment.

\begin{df}      \label{ready}
  Let $Z$ be a stage of a quantum circuit and let $G$ be a gate not in
  $Z$.  We say that $G$ is \emph{ready}
  at $Z$ if all its direct prerequisites (and therefore all its
  prerequisites) are in $Z$. In this case, $Z\cup\{G\}$ is also a
  stage, and we denote it by $Z+G$.
\end{df}

The idea behind this definition is that, after the gates in $Z$ have
fired, $G$ is ready to be fired next. Firing it would then bring the
computation to the stage $Z+G$.  There may, of course, be several
gates that are ready at $Z$, and any one (or more) of them could be
fired next.

Notice for future reference that, if a gate $G$ is ready at a stage
$Z$, then
\[
\text{Exit}(Z+G)=
\big(\text{Exit}(Z)-\pi(G)\big)
\sqcup \{(G,\mathrm{out},m):m\in o_G\},
\]
that is, firing $G$ after stage $Z$ adds to the exits the output ports
of $G$ and removes the producers that are providers for $G$. The
following proposition is just a reformulation of this observation in a
form that will be convenient later.

\begin{prop} \label{pass-gate}
  Let $Z$ be a stage, $G$ a gate that is ready at $Z$, and $R=\exx Z-\pi(G)$. Then
\[
\exx Z=R\sqcup \{\pi(G,\mathrm{in},l):l\in \iota_G\}
\]
and
\[
\exx{Z+G}=R\sqcup \{(G,\mathrm{out},m):m\in o_G\}.
\]
\end{prop}

We take advantage of this proposition to simplify some of our notation as follows.

\begin{notat} \label{identify}
Let $Z$, $G$, and $R=\exx Z-\pi(G)$ be
  as in Proposition~\ref{pass-gate}. We identify \exx Z with
  $R\sqcup \iota_G$ and thereby identify $Q^{\otimes\exx Z}$ with
  $Q^{\otimes R}\otimes Q^{\otimes \iota_G}$, using the bijection
  $l\mapsto \pi(G,\mathrm{in},l)$ for $l\in \iota_G$. Similarly, we
  identify $Q^{\otimes\exx{Z+G}}$ with
  $Q^{\otimes R}\otimes Q^{\otimes o_G}$, using the bijection
  $m\mapsto(G,\mathrm{out},m)$ for $m\in o_G$.  In other words, we
  omit mention of those two bijections and the maps they induce on the
  tensor powers of $Q$.
\end{notat}

It is also worth noting the two extreme cases of stages. The empty set
is a stage, the stage at which no gate has yet fired. Only the
circuit's input is available at this stage; formally,
$\text{Exit}(\emp)=I$.  The set of all gates is also a stage, the
stage after all the gates have fired. Its exits are the providers of the
output nodes.

The next theorem formalizes the idea that, once an input state
for the circuit is specified in $Q^{\otimes I}$, there is a
well-defined global state at each stage, where these global states for
different stages are related to each other by the action of the
gates.  In very abbreviated form, the theorem could be summarized as
saying that, given a circuit and an input state, there is a
well-defined computation of that circuit on that input. It is the
quantum analog of Theorem~\ref{b-thm}.

\begin{thm}     \label{q-thm}
Let a quantum circuit be given along with an input state vector
$\ket\psi\in Q^{\otimes I}$. Then there is a unique function assigning
to each stage $Z$ of the circuit a state vector $C(\ket\psi,Z)\in
  Q^{\otimes\exx Z}$ subject to the following requirements.
\begin{lsnum}
\item For the initial stage, we have $C(\ket\psi,\emp)=\ket\psi$.
\item If $G$ is ready at $Z$,then
\[
C(\ket\psi,Z+G)=(I_R\otimes U_G) C(\ket\psi,Z),
\]
where $R = \exx Z-\pi(G)$ is as in Proposition~\ref{pass-gate} and
$I_R$ is the identity operator on $Q^{\otimes R}$.
\end{lsnum}
\end{thm}

Requirement (2) in the theorem says, intuitively, that the gate $G$
acts on $C(\ket\psi,Z)$ to produce $C(\ket\psi,Z+G)$ by applying its
operator $U_G$ to the relevant part of this state, which, thanks to the
identifications in Notation~\ref{identify}, is $Q^{\otimes \iota_G}$. It
does nothing to the rest of $C(\ket\psi,Z)$, namely, the part in
$Q^{\otimes R}$.

\begin{pf}
Fix, for the whole proof, the circuit and the initial state $\ket\psi$.

To prove uniqueness of $C(\ket\psi,Z)$, we proceed by induction on the
cardinality of $Z$. If this cardinality is 0, then requirement~(1) in
the theorem ensures uniqueness of $C(\ket\psi,\emp)$.

Consider now a stage $Z$ that contains at least one gate. Because
$\prec^*$ is acyclic, $Z$ must contain a gate $G$ that is not a
prerequisite for any other gate in $Z$. Deletion of $G$ from our stage
thus produces another stage, which we call $Z'$ and which, by
induction hypothesis, has a uniquely defined $C(\ket\psi,Z')$. But
then $Z=Z'+G$, and $C(\ket\psi,Z)=C(\ket\psi,Z'+G)$ is uniquely
determined by requirement~(2) of the theorem. This completes the
induction step and thus completes the uniqueness proof.

It remains to prove the existence part of the theorem. The proof of
uniqueness given above implicitly provides a construction that almost
proves existence. Specifically, the uniqueness proof obtains
$C(\ket\psi,Z)$ by removing the gate $G$ to obtain $Z'$ with $Z=Z'+G$
and then acting by $I_R\otimes U_G$ on $C(\ket\psi,Z')$. This
$C(\ket\psi,Z')$ is, of course, obtained in the same way by removing a
gate $G'$ to get $Z'=Z''+G'$, and continuing in the same way until one
gets to the empty stage. Starting from $\ket\psi=C(\ket\psi,\emp)$, we
apply the gates (or rather their operators $I\otimes U$) in the
reverse of the order described above. That is, we have, using the
notation $R_i$ for $\exx{\{G_1,\dots,G_{i-1}\}}-\pi(G_i)$,
\[
C(\ket\psi,Z)=(I_{R_k}\otimes U_{G_k})\circ
(I_{R_{k-1}}\otimes U_{G_{k-1}})\circ\dots\circ
(I_{R_2}\otimes U_{G_2})\circ(I_{R_1}\otimes U_{G_1})\ket\psi
\]
for an enumeration $(G_1,G_2,\dots,G_k)$ of the gates in $Z$ that is
\emph{coherent} with $\prec$ in the sense that the prerequisites of
any gate appear earlier than the gate itself, i.e., if $G_i\prec G_j$
then $i<j$. (Intuitively, this enumeration is a sequentialization of
the circuit, in the sense that $G_1$ fires first, then $G_2$, etc.)
Note for future reference, that coherence with $\prec$ is the same as
coherence with the transitive closure $\prec^*$.

The issue that still needs to be addressed is that there are, in
general, several ways choose the gate $G$ in the induction step of the
uniqueness proof. We need that all enumerations $(G_1,G_2,\dots,G_k)$
of the gates in $Z$ coherent with $\prec$ produce the same
$C(\ket\psi,Z)$.

Once this is done, it will be clear that $C$ so defined satisfies the
requirements in the theorem. Indeed, requirement~(1) is immediate,
being the $k=0$ case of the definition where no operators act on
$\ket\psi$. To check requirement~(2), it suffices to apply the
definition with an arbitrary coherent enumeration for $Z$ and, for
$Z+G$, the enumeration obtained by appending $G$ as the last gate.

To show that any two coherent enumerations of $Z$ lead to the same
$C(\ket\psi,Z)$, we invoke Theorem~32 from \cite[\S4.2]{G242}:
If two
enumerations of a finite partially ordered set are both coherent with
the partial order, then one can be obtained from the other by a
sequence of interchanges of two consecutive elements, in such a way
that the enumerations at all steps of the process are coherent with
the partial order. We apply this result to the finite set $Z$ of gates
and the partial order $\prec^*$. We thus find that it suffices to
consider two enumerations that differ by interchanging just two
consecutive elements.

Suppose, therefore, that one enumeration is $(G_1,\dots,G_k)$ as above
and the other is obtained from it by interchanging $G_i$ and
$G_{i+1}$. These two enumerations give formulas for $C(\ket\psi,Z)$
that differ only in two of the factors of the form $I_R\otimes U_G$
that are being composed, and those two factors are adjacent. It is
tempting to say that we just need to prove that those factors commute,
but the situation is a bit more subtle because each of the two
relevant $R$'s depends on the preceding stages.

Let $Y$ be the stage just before either of the two critical gates
$G_i$ and $G_{i+1}$ acts, i.e., $Y=\{G_1,\dots,G_{i-1}\}$.
Then \exx Y
can be split into three disjoint subsets: the part $\pi(G_i)$ of providers
for the input ports of $G_i$, the analogous part $\pi(G_{i+1})$ for
$G_{i+1}$, and the rest $U$ of \exx Y. Formally, we observe that
$\pi(G_i)$ and $\pi(G_{i+1})$ are disjoint, because $\pi$ is
bijective, and we define $U=\exx Y-\pi(G_i)-\pi(G_{i+1})$.

Let us consider the action of the critical gates $G_i$ and $G_{i+1}$
with respect to the original enumeration
$(G_1,\dots,G_i,G_{i+1},\dots,G_k)$ and, in particular, let us look at
the $R_i$ and $R_{i+1}$ at those stages. In our formula for
$C(\ket\psi,Z)$, we determined $R_i$ by referring to
Proposition~\ref{pass-gate} with the gate $G_i$ and the stage just
before the action of $G_i$. In our present context, that stage is what
we are calling $Y$, and $R_i$ is therefore
$\exx Y-\pi(G_i)=\pi(G_{i+1})\sqcup U$.  Then $R_{i+1}$ is determined
by again referring to Proposition~\ref{pass-gate} but now with the
gate $G_{i+1}$ and the stage $Y+G_i$. So $R_{i+1}$ is
$\exx {Y+G_i}-\pi(G_{i+1})$ This $R_{i+1}$ could, a priori, differ
from $\exx Y-\pi(G_{i+1})=\pi(G_i)\sqcup U$, because we now have
\exx{Y+G_i} rather than \exx Y. Fortunately, there is no real
difference. To see this, first consult Proposition~\ref{pass-gate} to
see that \exx{Y+G_i} differs from \exx Y only by (1)~removal of ports
that are in $\pi(G_i)$ and (2) addition of output nodes of $G_i$. The
removals in (1) make no difference for us, because $\pi(G_i)$ is disjoint
from $\pi(G_{i+1})$. The additions in (2) also make no difference for the
following, less trivial reason. Recall that the enumeration with $G_i$
and $G_{i+1}$ interchanged is also coherent with $\prec$. So we know
that $G_i\not\prec G_{i+1}$; no output node of $G_i$ can be the
provider for an input node of $G_{i+1}$. And this is just what we need
to ensure that the additions (2) don't matter.

Thus, for the enumeration $(G_1,\dots,G_i,G_{i+1},\dots,G_k)$, we have
$R_i=\exx Y-\pi(G_i)=\pi(G_{i+1})\sqcup U$ and
$R_{i+1}=\exx Y-\pi(G_{i+1})=\pi(G_i)\sqcup U$.  An exactly analogous
argument gives the same $R$'s for the alternative ordering
$(G_1,\dots,G_{i+1},\allowbreak G_i,\dots,G_k)$. That is, the same
two operators $I_{R_i}\otimes U_{G_i}$ and
$I_{R_{i+1}}\otimes U_{G_{i+1}}$ occur in both versions of the formula
for $C$ but in reversed order. So all we still need to prove is that
these two operators commute. Fortunately, that is easy. The first is
$I_U\otimes U_{G_i}\otimes I_{\pi(G_{i+1})}$ and the second is
$I_U\otimes I_{\pi(G_{i})} \otimes U_{G_{i+1}}$. These commute
because, in each of the three tensor factors, at least one of them is
the identity operator.  This completes the proof that $C$ is
well-defined and thus completes the proof of the theorem.
\end{pf}

The following corollary records information implicit in the proof of
Theorem~\ref{q-thm}, namely that $C(\ket\psi,Z)$ is, for each fixed
$Z$, obtained from $\ket\psi$ in a uniform and unitary manner.

\begin{cor}
  For every quantum circuit and every stage $Z$, there is a unitary
  operator $Q^{\otimes I}\to Q^{\otimes\exx Z}$ sending each input
  state vector $\ket\psi$ to the $C(\ket\psi,Z)$ described in
  Theorem~\ref{q-thm}.
\end{cor}

\begin{pf}
  The desired unitary operator is the operator
\[
(I_{R_k}\otimes U_{G_k})\circ
(I_{R_{k-1}}\otimes U_{G_{k-1}})\circ\dots\circ
(I_{R_2}\otimes U_{G_2})\circ(I_{R_1}\otimes U_{G_1}),
\]
used in the proof of Theorem~\ref{q-thm} and shown there to be
independent of the sequentialization of the circuit.
\end{pf}

Given a quantum circuit and an input state $\ket\psi\in Q^{\otimes I}$,
Theorem~\ref{q-thm} provides a complete description of the resulting
computation. That includes, in particular, the final result of these
computations, namely $C(\ket\psi,Z)$ where $Z$ is the set of all the
gates in the circuit. But it also includes intermediate results
$C(\ket\psi,Z)$ for all stages $Z$, and these tell what happens,
step-by-step, in any sequentialization of the
computation. Furthermore, even if one does not fully sequentialize the
computation but allows several gates to act simultaneously, then,
first, any simultaneously acting gates must be incomparable under
$\prec^*$ because a gate can act only after its inputs have been
produced by its prerequisite gates, and, second, after any part of
such a computation has taken place, the gates in that part constitute
a stage. Theorem~\ref{q-thm} thus provides a well-defined state after
that part of the computation.

\begin{rmk}
  Theorems~\ref{b-thm} and \ref{q-thm} say, in the Boolean and quantum
  cases respectively, that a circuit and initial data produce a
  well-defined computation, but the detailed formulations differ
  significantly. Theorem~\ref{b-thm} provides a separate bit $C(x)$
  for every node $x$. We have already explained that it is
  unreasonable to expect an exact analog for quantum circuits,
  providing a separate qubit for every node, because the qubits can be
  entangled.  This is why Theorem~\ref{q-thm} assigns quantum states
  not to individual nodes but to (the exits of) whole stages.

  In the opposite direction, though, we can easily obtain a Boolean
  analog of the quantum result. Observe that the definitions of
  ``stage'', ``exit'', ``ready'', and ``$Z+G$'' can be applied
  verbatim to Boolean circuits. Furthermore,
  Proposition~\ref{pass-gate} remains correct and we can make
  identifications like those in Notation~\ref{identify} for products
  of sets instead of tensor products of vector spaces.  In light of
  these observations, we can transcribe Theorem~\ref{q-thm} to the
  Boolean situation, obtaining the following corollary of
  Theorem~\ref{b-thm}.
\end{rmk}

\begin{cor}
Let a Boolean circuit be given along with an input
$a\in \{0,1\}^{ I}$. Then there is a unique function assigning
to each stage $Z$ of the circuit a state $C(a,Z)\in
  \{0,1\}^{\exx Z}$ subject to the following requirements.
\begin{lsnum}
\item For the initial stage, we have $C(a,\emp)=a$.
\item If $G$ is ready at $Z$, then
\[
C(a,Z+G)=(I_R\times g_G) C(a,Z),
\]
where $R = \exx Z-\pi(G)$ is as in Proposition~\ref{pass-gate} and $I_R$
is the identity function on $\{0,1\}^{ R}$.
\end{lsnum}
\end{cor}

\begin{pf}
The desired $C(a,Z)$ in this corollary is the function assigning to
each $x\in\exx Z$ the bit $C(x)$ from Theorem~\ref{b-thm}. The
required properties of $C$ in the present corollary follow easily
from the properties of the earlier $C$ in Theorem~\ref{b-thm}.
\end{pf}

\end{document}